\documentclass[pre,twocolumn]{revtex4}
\usepackage{graphicx}
\usepackage{amssymb,amsmath,enumerate}

\newcommand{\be}{\begin{eqnarray}}
\newcommand{\ee}{\end{eqnarray}}
\newcommand{\D}{\mathrm{d}}

\begin{document}

\title{Rectification of asymmetric surface vibrations with dry friction: an exactly solvable model}

\author{A. Baule$^1$ and P. Sollich$^2$ }

\affiliation{
$^1$School of Mathematical Sciences, Queen Mary University of London, Mile End Road, London E1 4NS, UK\\
$^2$Department of Mathematics, King's College London, London WC2R 2LS, UK
}

\begin{abstract}

We consider a stochastic model for the directed motion of a solid object due to the rectification of asymmetric surface vibrations with Poissonian shot-noise statistics. The friction between the object and the surface is given by a piecewise-linear friction force. This models the combined effect of dynamic friction and singular dry friction. We derive an exact solution of the stationary Kolmogorov-Feller (KF) equation in the case of two-sided exponentially distributed amplitudes. The stationary density of the velocity exhibits singular features such as a discontinuity and a delta-peak singularity at zero velocity, and also contains contributions from non-integrable solutions of the KF equation. The mean velocity in our model generally varies non-monotonically as the strength of the dry friction is increased, indicating that transport improves for increased dissipation.

\end{abstract}

\date{\today}

\maketitle

\section{Introduction}

Understanding noise-induced transport on micro- and nanoscales in the absence of an external force gradient has been an active research area in recent years. In particular, it has led to the notion of {\it Brownian motors}, which are used to model, e.g.\ the propulsion mechanisms of cellular motor proteins. Here, directed motion results quite generically from the combination of (i) \textit{non-equilibrium energy} input, e.g.\ due to chemical or mechanical fluctuations, and (ii) a spatial or dynamical \textit{asymmetry}. A large variety of Brownian motor models implementing these principles have been developed, where typically the fluctuations are rectified using a ratchet-shaped potential; see the review articles \cite{Reimann02,Haenggi09} and references therein.

In related work, small objects such as hydrogels, droplets and micron sized particles have been moved over solid surfaces using asymmetric surface oscillations or noisy vibrations without a net force \cite{Daniel02,Daniel04,Mahadevan04,Daniel05,Eglin06,Buguin06,Fleishman07}. This kind of noise-induced transport relies on a ratchet effect in the velocity coordinate, which requires in addition to (i) and (ii) a {\em nonlinear resistance} of the moving object to the imposed force. Such a nonlinearity can originate, e.g.\ from nonlinear friction or wetting hysteresis. In the case of a solid object moving over a solid surface, the nonlinear dry (Coulombic) friction between the two solids leads to hysteresis of the contact line, which effectively rectifies the fluctuations: dry friction makes the object stick to the surface so that a threshold force is needed to move it. Directed motion results when the forces do not overcome the threshold symmetrically, and this may happen even though the total force is zero on average.

A simple model for noise-induced transport can thus be implemented by coupling asymmetric fluctuations to an object subject to nonlinear friction. In Ref.~\cite{Cebiroglu10}, shot-noise fluctuations given by a two-state (dichotomous) Markov process were investigated with different types of nonlinear friction. In our recent letter \cite{Baule12}, we considered singular dry friction in combination with one-sided Poissonian shot noise (PSN). PSN can be considered as a standard model for a mechanical random force and generalizes the usual Gaussian noise, to which it converges in a certain limit. Since PSN contains an intrinsic timescale (the mean waiting time between successive shots) and also allows for a variation of the shot amplitudes, it is a suitable model for a broad range of different fluctuation scenarios.

The interplay of dry friction and PSN not only leads to a non-zero mean velocity of the object in the absence of a mean force input, but also to non-monotonic and singular features in the transport properties, as discussed in Ref.~\cite{Baule12}. In the present work, we generalize these results and obtain in particular an exact solution of the Kolmogorov-Feller (KF) equation for the stationary velocity distribution for the case where the PSN is two-sided with exponentially distributed amplitudes. In this case, non-integrable solutions of the KF equation contribute to the stationary distribution in a certain parameter regime. The mean velocity as a function of the dry friction strength varies non-monotonically as the dry friction is increased, reaching a maximum at a particular dry friction strength. The transport in our model therefore generally improves for larger friction, over a considerable range; this is highly counterintuitive.

The paper is organized as follows. In Sec.~\ref{Sec_model} we introduce our model for the directed motion of an object moving under the influence of dry friction and PSN. General results on the form of the stationary velocity distribution are obtained. In Sec.~\ref{Sec_exp} we derive the explicit solution for PSN with a two-sided exponential amplitude distribution. Singular features as well as contributions from non-integrable solutions of the KF equation are discussed. In Sec.~\ref{Sec_trans} we investigate the non-monotonic behaviour of the transport properties of our model. Possible generalizations of our results and their relevance to experiments are discussed in Sec.~\ref{Sec_dis}.

\section{Directed motion due to dry friction and PSN}
\label{Sec_model}

We consider the dynamics of a solid object moving over a solid surface subject to friction and noise $\xi(t)$. In one dimension, to which we restrict ourselves for simplicity, and if we set the mass of the object to $m=1$, its velocity satisfies the equation of motion
\be
\label{model}
\dot{v}(t)=-\frac{1}{\tau}v(t)-\sigma(v)\Delta+\xi(t)-a.
\ee
The friction between the two solids is expressed phenomenologically by two friction terms: a {\it dynamic friction} linear in $v$ with strength $1/\tau$, where $\tau$ is the inertial relaxation time, and a {\it dry friction} $-\sigma(v)\Delta$ with strength $\Delta$. The sign function $\sigma(v)$ is defined as
\be
\sigma(v)=\left\{\begin{array}{ll} 1,\qquad & v< 0 \\ & \\ 0, & v=0\\ & \\ -1, & v<0.\end{array}\right.
\ee
The singularity in $\sigma(v)$ at $v=0$ models the non-analytic behaviour observed in the dynamics of a solid object due to dry friction. In particular, the dry friction force in Eq.~(\ref{model}) can lead to complicated nonlinear stick-slip dynamics, where the object alternates in an oscillatory or chaotic way between sticking and sliding states \cite{Persson,Urbakh04}.

The force $a$ denotes a constant drift force. We take this as the mean of the noise $\xi(t)$, setting $a=\left< \xi(t)\right>$ so that overall there is no net force acting on the object. Without loss of generality we only consider $a\ge 0$, since the case of negative $a$ follows by symmetry. Despite the absence of a net force bias, the object exhibits a non-zero mean velocity due to the rectification of the asymmetric noise. These transport properties will be discussed in more detail in Sec.~\ref{Sec_trans} below.

If we write $F(v)$ for the sum of all the deterministic forces, Eq.~(\ref{model}) can be written in the generic form
\be
\label{model2}
\dot{v}(t)=F(v)+\xi(t),
\ee
where $F(v)$ denotes a piecewise-linear force with a discontinuity at $v=0$:
\be
\label{force2}
F(v)=\left\{\begin{array}{l l}F_+(v),\qquad & v>0\\ & \\ F_-(v),\qquad & v<0.\end{array}\right.
\ee
The two branches of the force are given by
\be
\label{forcepm}
F_\pm(v)&=&-\frac{1}{\tau}(v-v_\pm),
\ee
where we define the two deterministic fixed points $v_\pm$ from the conditions $F_\pm(v_\pm)=0$,
\be
\label{fixp}
v_\pm=-(a\pm\Delta)\tau.
\ee
Below, in Sec.~\ref{Sec_smooth}, we consider a regularized version of $F(v)$ where the discontinuity at $v=0$ is replaced by a linear interpolation over a small velocity range.

We consider Poissonian shot noise (PSN) as a model for the noise $\xi(t)$ in Eq.~(\ref{model}). PSN is a mechanical random force that can be considered a generalization of the usual Gaussian white noise. It can be represented by a sequence of delta shaped pulses with random amplitudes $A$ \cite{Feynman}
\begin{eqnarray}
\label{PSN}
z(t)=\sum_{k=1}^{n}A_k\delta(t-t_k).
\end{eqnarray}
The waiting time between the successive pulses at times $t_k$ is assumed to be exponentially distributed with parameter $\lambda$, where $\lambda$ is the rate at which pulses arrive. Then $n$, the number of pulses in time $t$, follows a Poisson distribution
\begin{eqnarray}
P(n)=\frac{(\lambda t)^n}{n!} e^{-\lambda t},
\end{eqnarray}
with mean $\lambda t$. When a pulse occurs, its amplitude $A_k$ is sampled, independently for each pulse, from a distribution $\rho(A)$. The mean and the covariance of $z(t)$ are then given by
\be
\left<z(t)\right>&=&\lambda\left<A\right>_\rho\\
\left<z(t)z(t')\right>-\left<z(t)\right>\left< z(t')\right>&=&\lambda\left<A^2\right>_\rho \delta(t-t'),
\ee
where $\left<...\right>_\rho$ indicates an average over the amplitude distribution $\rho(A)$. In the limit of vanishing pulse amplitudes $A\to 0$ and a divergent pulse frequency $\lambda\to \infty$, with $\left<A\right>_\rho=0$ and $\lambda\left<A^2\right>_\rho=\mbox{const.}=2D$, PSN reduces to Gaussian white noise. 

In Eq.~(\ref{model}) we choose $\xi(t)$ as a superposition of two PSN processes
\be
\xi(t)=z_1(t)+z_2(t),
\ee
where $z_1(t)$ and $z_2(t)$ are specified by the pulse rates $\lambda_{1,2}$ and amplitude distributions $\rho_{1,2}(A)$. In order to have a noise with zero mean we set the constant drift force $a$ to
\be
a=\langle\xi(t)\rangle=\lambda_1\left<A\right>_{\rho_1}+\lambda_2\left<A\right>_{\rho_2},
\ee
where the mean amplitudes are given by
\be
\left<A\right>_{\rho_j}=\int_{-\infty}^\infty A\,\rho_j(A)\,\D A,\qquad j=1,2.
\ee

In the Gaussian limit of the PSN, Eq.~(\ref{model2}) converges to Brownian motion with dry friction, which has recently been investigated both theoretically and experimentally \cite{deGennes05,Hayakawa04,Baule10,Baule11,Touchette10,Menzel10,Goohpattader09,Goohpattader10,Mettu10}. No directed motion results in this case unless an additional bias, such as a constant force, acts on the object. On the other hand, asymmetric PSN has been shown to induce directed transport for overdamped dynamics in a symmetric potential \cite{Luczka95}. The effect of a two-state (dichotomous) shot noise was investigated in Ref.~\cite{Cebiroglu10} for different types of nonlinear friction, but no singular features were reported. The particular case of one-sided PSN, discussed in Ref.~\cite{Baule12}, is obtained from the present model by setting $\lambda_2=0$.

The time dependent velocity distribution $p(v,t)$ obeys the Kolmogorov-Feller equation \cite{Feller,Haenggi78}
\be
\label{KF}
\frac{\partial}{\partial t}p(v,t)&=&-\frac{\partial}{\partial v}F(v)p(v,t)\nonumber\\
&&+\sum_{j=1}^2\lambda_j\left(\left< p(v-A,t)\right>_{\rho_j}-p(v,t)\right),
\ee
where the average $\left< p(v-A,t)\right>_{\rho_j}$ indicates the convolution
\be
\label{nonlocal}
\left< p(v-A,t)\right>_{\rho_j}&=&\int_{-\infty}^\infty p(v-A,t)\rho_j(A)\,\D A.
\ee
We are interested in the properties of the stationary probability distribution $p(v)$. In the Gaussian limit, Eq.~(\ref{KF}) becomes the Fokker-Planck equation with the piecewise-linear force $F(v)$,
\be
\label{FP}
\frac{\partial}{\partial t}p(v,t)&=&-\frac{\partial}{\partial v}F(v)p(v,t)+D\frac{\partial^2}{\partial v^2} p(v,t).
\ee
Under the conditions of stationarity and a zero probability current, the stationary distribution $p(v)$ is then obtained from
\be
\label{FPstat}
F(v)p(v)=D\frac{\D}{\D v} p(v),
\ee
which can be solved by integration. Eq.~(\ref{FPstat}) indicates that $p(v)$ is continuous even in the presence of the discontinuous force: the discontinuity at $v=0$ on the left-hand side of Eq.~(\ref{FPstat}) can be balanced by a discontinuous derivative on the right-hand side, so that the stationary distribution has a cusp singularity at the point where the force is discontinuous.

In the following we focus on the non-Gaussian parameter regime of the PSN. Here, the stationarity condition becomes, instead of Eq.~(\ref{FPstat}),
\be
\label{zerocurr}
F(v)p(v)=\int_{-\infty}^\infty G(v-v')p(v')\D v',
\ee
where the diffusion operator is expressed in terms of a Green's function (see below). The integral equation (\ref{zerocurr}) indicates that PSN induces a {\it non-local} diffusion of the degree of freedom, in contrast to the local normal diffusion that is generated by standard Gaussian white noise. Eq.~(\ref{zerocurr}) can be derived in a straightforward matter by performing a Fourier transformation (FT) of Eq.~(\ref{KF}) with $\dot{p}=0$. In Fourier space, where we denote Fourier transforms by a $\sim$, the Green's function is given by
\be
\label{Greens1}
\tilde{G}(k)=i\sum_{j=1}^2\lambda_j\left(\frac{\tilde{\rho}_j(k)-1}{k}\right).
\ee
This leads to the representation
\be
\label{Greens2}
G(v)&=&\sum_{j=1}^2\lambda_j\int^v_{-\infty}\rho_j(v')\D v'-(\lambda_1+\lambda_2)\theta(v).
\ee
This expression can also be obtained directly by integration of Eq.~(\ref{zerocurr}); $\theta(v)=(1+\sigma(v))/2$ indicates the Heaviside step function. Because of this contribution the Green's function always has a jump discontinuity at $v=0$, even if the amplitude distributions $\rho_j(A)$ are smooth. 

Let us now investigate the properties of $p(v)$ at the discontinuity of the force. We assume that the stationary distribution has the form
\be
\label{pdist1}
p(v)=\left\{\begin{array}{l l} c_1p_+(v),\qquad & v>0\\ & \\ c_2p_-(v),\qquad & v<0,\end{array}\right.
\ee
where the constants $c_1$ and $c_2$ guarantee normalization. Substituting $v= 0^\pm$ into Eq.~(\ref{zerocurr}) yields
\be
\label{zb1}
\frac{F_+(0^+)}{F_-(0^-)}=\frac{c_2p_-(0^-)}{c_1p_+(0^+)}.
\ee
because the convolution integral on the r.h.s. is continuous at $v=0$. 
The last result together with Eq.~(\ref{force2}) implies that $p(v)$ must have a discontinuity at $v=0$. Once the functions $p_\pm(v)$ have been found, Eq.~(\ref{zb1}) together with the normalization constraint $\int p(v)\D v=1$ gives two equations for the two constants $c_1,c_2$.

However, the above reasoning can be applied only when $F_+(0^+)$ and $F_-(0^-)$ have the same sign. When they do not, Eq.~(\ref{zb1}) clearly cannot be satisfied. In this case, there has to be a finite change in probability mass at the crossover from $v<0$ to $v>0$, which can only stem from a delta peak at $v=0$. We therefore need a more general ansatz
\be
\label{pdist2}
p(v)=\left\{\begin{array}{l l} c_1p_+(v),\qquad & v>0\\ & \\ \Gamma_0\delta(v),\qquad & v=0\\ & \\ c_2p_-(v), \qquad& v<0.\end{array}\right.
\ee
The r.h.s.\ of Eq.~(\ref{zerocurr}) now has a jump discontinuity $-(\lambda_1+\lambda_2)\Gamma_0 \theta(v)$ from the convolution of the step function in $G$ with the delta peak in $p(v)$. Other than this the r.h.s.\ is still continuous, so comparing values at $v=0^+$ and $v=0^-$ in Eq.~(\ref{zerocurr}) gives
\be
\label{zb2}
F_+(0^+)c_1p_+(0^+)=F_-(0^-)c_2p_-(0^-)-\bar{\lambda}\Gamma_0,
\ee
where we have abbreviated $\bar{\lambda}=\lambda_1+\lambda_2$. When $\Gamma_0=0$ this reduces to Eq.~(\ref{zb1}) as it should.

In order to determine the three unknowns $c_1,c_2,\Gamma_0$ we now require one further condition in addition to Eq.~(\ref{zb2}) and the overall normalization condition. This condition can be obtained from the derivative of the stationarity condition Eq.~(\ref{zerocurr})
\be
&&F'(v)p(v)+F(v)p'(v)\nonumber\\
&&=\sum_{j=1}^2\lambda_j\int_{-\infty}^\infty \rho_j(v-v')p(v')\D v'-\bar{\lambda} p(v),
\ee
which also follows directly from Eq.~(\ref{KF}). Again comparing terms across the discontinuity leads to
\be
\label{zb3}
&&c_1\left(F'_+(0^+)p_+(0^+)+F_+(0^+)p'_+(0^+)+\bar{\lambda} p_+(0^+)\right)\nonumber\\
&&=c_2\left(F'_-(0^-)p_-(0^-)+F_-(0^-)p'_-(0^-)+\bar{\lambda}p_-(0^-)\right)\nonumber\\
&&\quad+ \Gamma_0\sum_{j=1}^2\left(\lambda_j\rho_j(0^+)-\lambda_j\rho_j(0^-)\right),
\ee
and this provides the third condition on the unknown normalization constants. The main problem is thus to determine the solutions $p_+(v)$ and $p_-(v)$ on either side of the discontinuity. This would in principle require solving the integral equation (\ref{zerocurr}), which is a difficult task. In the following we pursue an alternative solution method: from Eq.~(\ref{zerocurr}) we derive a local ordinary differential equation that can be solved separately for $v<0$ and $v>0$. This approach is feasible for amplitude distributions $\rho_j$ that are given as single exponentials, as we demonstrate in the next section.

\section{Exact solution for exponential amplitude distributions}
\label{Sec_exp}

Exact expressions for the stationary distributions $p_\pm(v)$ can be derived if the amplitude distributions are taken as single exponentials
\be
\label{2exp}
\rho_1(A)=\frac{\Theta(A)}{A_1}e^{-A/A_1},\qquad\rho_2=\frac{\Theta(-A)}{A_2}e^{A/A_2},
\ee
so that the noise in Eq.~(\ref{model}) is two-sided PSN. Here the parameters $A_1,A_2>0$ relate to the mean amplitudes defined above as $\left<A\right>_{\rho_1}=A_1$ and $\left<A\right>_{\rho_2}=-A_2$, and the mean of the two-sided PSN is
\be
\label{a_def}
a=\lambda_1 A_1 - \lambda_2 A_2.
\ee
Eq.~(\ref{2exp}) leads to the FTs
\be
\tilde{\rho}_1(k)=\frac{1}{1-iA_1 k},\qquad\tilde{\rho}_2(k)=\frac{1}{1+iA_2k}.
\ee
Using Eq.~(\ref{Greens1}) the Greens function $\tilde{G}(k)$ is in this case
\be
\label{Gexp}
\tilde{G}(k)=-\frac{\lambda_1A_1}{1-iA_1k}+\frac{\lambda_2A_2}{1+iA_2 k}.
\ee 
Multiplying the FT of Eq.~(\ref{zerocurr}) by $(1+iA_2k)(1-iA_1k)$ one thus obtains an ordinary differential equation of second order
\be
\label{ODE}
&&\left(1+A_1\frac{\D}{\D v}\right)\left(1-A_2\frac{\D}{\D v}\right)F(v)p(v)\nonumber\\
&&=-a\, p(v)+\bar{\lambda} A_1A_2\frac{\D}{\D v}p(v).
\ee 
This ODE can be solved separately for $v>0$ and $v<0$, and since $F(v)$ is linear in each range, these solutions can be found using FTs. These solutions are identical to the ones obtained directly from a FT of Eq.~(\ref{zerocurr}) under the assumption of a purely linear force $F(v)=F_\pm(v)$. We note, however, that for general amplitude distributions solving Eq.~(\ref{zerocurr}) in this way would not yield the correct solution. Instead one would need to derive and solve the corresponding local differential equation. For the exponential distributions considered here, the solution is then 
\be
\label{ppm}
p_\pm(v)&=&f(v-v_\pm),
\ee
where  the two fixed points $v_\pm$ are defined in Eq.~(\ref{fixp}). In terms of the shifted velocity $\tilde{v}=v-v_\pm$, $f(\tilde{v})$ is the solution of Eq.~(\ref{ODE}) with the linear force $F(\tilde{v})=-\tilde{v}/\tau$:
\be
f(\tilde{v})&=&\int_{-\infty}^\infty\exp\left\{-ik\tilde{v}-i\tau\int^k\tilde{G}(k')\D k'\right\}\D k,\nonumber\\
&=&\int_{-\infty}^\infty\exp\left\{-ik\tilde{v}\right\}(1-iA_1k)^{-\lambda_1\tau}\times\nonumber\\
&&(1+iA_2k)^{-\lambda_2\tau}\D k.
\ee

For the following it is convenient to define
\be
f_1^\pm(\tilde{v})&=&\Theta(\pm \tilde{v})(\pm \tilde{v})^{\lambda_1\tau-1}e^{-\tilde{v}/A_1}\\
f_2^\pm(\tilde{v})&=&\Theta(\pm \tilde{v})(\pm \tilde{v})^{\lambda_2\tau-1}e^{\tilde{v}/A_2},
\ee
so that, if we discard irrelevant proportionality constants,
\be
\label{FTsol}
f(\tilde{v})&=&f_1^+(\tilde{v})*f_2^-(\tilde{v}).
\ee
The solutions $p_\pm(v)$ follow by mapping back to the original velocity via $\tilde{v}=v-v_\pm$. They are thus given as the convolution (indicated by ``$*$" above) of two Gamma distributions, shifted by $v_\pm$. Both $p_\pm(v)$ are normalizable ({\em integrable}) on the respective domains $(-\infty,0)$ and $(0,\infty)$, which is a consequence of the use of FTs.

However, Eq.~(\ref{FTsol}) does not provide the complete solution of Eq.~(\ref{ODE}). We can also find {\em non-integrable} (over the positive or negative real axis) solutions of Eq.~(\ref{ODE}) if we use the Laplace transformation (LT). These solutions can contribute to the stationary density because of the piecewise-linearity of $F(v)$, as discussed in more detail below. Similar to the FT case, the LT solutions of Eq.~(\ref{ODE}) will be given as $p_\pm(v)=h(v-v_\pm)$, with $h(\tilde{v})$ the solution of Eq.~(\ref{ODE}) for $F(\tilde{v})=-\tilde{v}/\tau$ as before. Solving Eq.~(\ref{ODE}) for $\tilde{v}>0$ with the condition $h(0)=0$ using LTs, one obtains in a straightforward way (denoting LTs with a hat)
\be
\hat{h}(s)=(1+A_1s)^{-\lambda_1\tau}(A_2s-1)^{-\lambda_2\tau},
\ee
which yields upon inversion
\be
\label{LTsol}
h(\tilde{v})=f_1^+(\tilde{v})*f_2^+(\tilde{v}).
\ee
Likewise one finds for $\tilde{v}<0$
\be
\label{LTsol2}
h(\tilde{v})=f_1^-(\tilde{v})*f_2^-(\tilde{v}).
\ee
Setting $\tilde{v}=v-v_\pm$ leads to the LT solutions of Eq.~(\ref{ODE}) for the piecewise-smooth force Eq.~(\ref{force2}).

Here, it is crucial to distinguish between the cases $\Delta>a$ and $\Delta<a$, which are equivalent to $F_+(0^+)F_-(0^-)<0$ and $F_+(0^+)F_-(0^-)>0$, respectively (cf.\ Eq.~(\ref{forcepm})). In the case $\Delta>a$ we have $v_+<0$ and $v_->0$, so that the two LT solutions are given by 
\be
\label{p1}
p_1(v)&=&\Theta(v)h(v-v_+)\\
p_2(v)&=&\Theta(-v)h(v-v_-).
\ee
However, neither can contribute to the stationary density since they diverge for $v\to\pm\infty$. In this case the solution is thus given by Eq.~(\ref{pdist2}) with the $p_\pm(v)$ of Eq.~(\ref{ppm}). It contains the three undetermined constants $c_1$, $c_2$, $\Gamma_0$, which can be found from the normalization constraint and the consequences Eqs.~(\ref{zb2}) and (\ref{zb3}) of the stationarity condition.

In the case $\Delta<a$, on the other hand, we have both $v_+<0$ and $v_-<0$. This yields the solution $p_1(v)$, Eq.~(\ref{p1}), and in addition
\be
p_3(v)&=&\Theta(-v)\Theta(v_--v)h(v-v_-)\\
p_4(v)&=&\Theta(-v)\Theta(v-v_-)h(v-v_-).
\ee
The solutions $p_1(v)$ and $p_3(v)$ diverge for $v\to\pm\infty$ and have to be discarded. Only $p_4(v)$ is normalizable on the domain $(-\infty,0)$ and can therefore contribute to the overall solution for $v<0$. In the case $\Delta<a$ one thus obtains the stationary distribution in the form~(\ref{pdist1}) with
\be
\label{pplus}
p_+(v)=f(v-v_+)
\ee
as before, and
\be
\label{pneg}
p_-(v)&=&f(v-v_-)+c_3\Theta(v-v_-)h(v-v_-),
\ee
A separate normalization constant $c_3$ appears here for the LT contribution. Because of this, an extra constraint is needed to fix all of $c_1$, $c_2$, $c_3$, beyond normalization and Eq.~(\ref{zb1}). This is again provided by Eq.~(\ref{zb3}), but with $\Gamma_0=0$.

We note finally that $f(v)$ and $h(v)$ only exhaust three out of the four possible convolutions of the form $f_1^\pm(\tilde v)*f_2^\pm(\tilde{v})$. The fourth combination $f_1^-(\tilde{v})*f_2^+(\tilde{v})$ does not appear above because it is undefined: the integral defining the convolution always diverges.

\subsection{Linearly smoothed force}
\label{Sec_smooth}

In order to understand in more detail the emergence of the delta-peak for $\Delta>a$ for the discontinuous force $F(v)$, it is helpful to consider a linearly smoothed representation of the sign-function
\be
\label{sigeps}
\sigma_\epsilon(v)=\left\{\begin{array}{ll} 1,\qquad & v\ge\epsilon \\ & \\ v/\epsilon,\qquad & -\epsilon<v<\epsilon\\ & \\ -1,\qquad & v\le -\epsilon\end{array}\right.
\ee
which obeys
\be
\sigma(v)=\lim_{\epsilon\to 0}\sigma_\epsilon(v).
\ee
In the linearized region, the force $F(v)$, Eq.~(\ref{force2}), is thus
\be
F_\epsilon(v)=-\frac{\Delta}{\epsilon}\left(v+\frac{\epsilon}{\Delta}a\right),\qquad -\epsilon<v<\epsilon,
\ee
for small $\epsilon$. Performing the above derivation of $p_\pm(v)$ using FTs with this linear force leads to the stationary distribution
\be
\label{peps}
p_\epsilon(v)\propto\left|\frac{v}{\epsilon}+\frac{a}{\Delta}\right|^{\bar{\lambda}\epsilon/\Delta-1},
\ee
for $-\epsilon<v<\epsilon$ and small $\epsilon$. One can see that $p_\epsilon(v)$ has a divergence at $v=-a\epsilon/\Delta$ and this lies inside the smoothing region for $\Delta > a$. At the same time, $p_\epsilon(v)$ has finite probability mass when integrated over $[-\epsilon,\epsilon]$. It therefore turns into a delta-peak in the limit $\epsilon\to 0$ of a genuinely discontinuous force.

\section{Features of the stationary density}
\label{Sec_pdens}

We now explore the properties of the stationary velocity distribution $p(v)$ in the various parameter regimes. For definiteness we will set the time scale $\tau=1$ and the velocity scale $A_2=1$. We also fix $a\ge 0$ so that $\lambda_1A_1\ge \lambda_2 A_2$; the case $a<0$ follows by symmetry. One of our more surprising findings will be that for both $\Delta > a$ and $\Delta< a$ the stationary velocity distribution can exhibit characteristic double peaks. 

One important feature of $p(v)$ is that the building blocks $f(\tilde{v})$ and $h(\tilde{v})$, defined in Eqs.~(\ref{FTsol}), (\ref{LTsol}) and~(\ref{LTsol2}), 
depend only on the rate and amplitude parameters $\lambda_1$, $\lambda_2$ and $A_1$. The friction strength $\Delta$ only appears in the way these functions are shifted, via $\tilde{v}=v-v_\pm$, and in the conditions for the coefficients $c_1$, $c_2$ and $c_3$ or $\Gamma_0$ with which they are weighted in $p(v)$. It therefore makes sense to consider first these more elementary functions, and in particular $f(\tilde{v})$ from  Eq.~(\ref{FTsol}). This function always features in $p(v)$ while $h(\tilde{v})$ appears only for $\Delta < a$.

\begin{figure}
\begin{center}
\includegraphics[width=8cm]{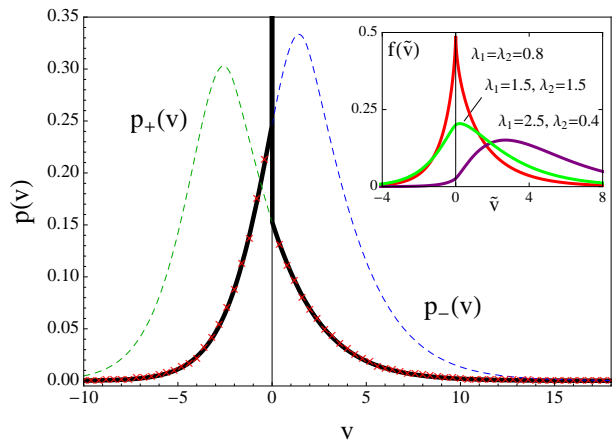}
\caption{\label{Fig_pdens1}(Color online) The stationary density $p(v)$ for $\Delta>a$, Eq.~(\ref{pdist2}), for 
parameter values $\lambda_1=1.5$, $\lambda_2=3$, $A_1=2$, $\Delta=2$. The partial densities $p_\pm(v)=f(v-v_\pm)$ are indicated by dashed lines.   The red crosses are results from a direct numerical simulation of Eq.~(\ref{model}) using an increment method for PSN \cite{Kim07}. Inset: The function $f(\tilde{v})$, Eq.~(\ref{FTsol}), for $A_1=2$ and rates $(\lambda_1,\lambda_2)$ as indicated. For $\lambda_1+\lambda_2<1$, $f(\tilde{v})$ has a power law peak at the origin.}
\end{center}
\end{figure}

If we go back to how $f(\tilde{v})$ was calculated above, we see that it represents the stationary velocity distribution of a particle subject to a linear restoring force $-\tilde{v}/\tau$ and two-sided PSN whose mean has not been subtracted off. Three different plots of $f(\tilde{v})$ are shown in the inset of Fig.~\ref{Fig_pdens1} to illustrate the range of possible behaviours. One sees that $f(\tilde{v})$ is skewed, and this is generically the case unless $\lambda_1=\lambda_2$ and $A_1=A_2$.

When both noise rates are larger than the inverse inertial relaxation time, $\lambda_1,\lambda_2>1/\tau \equiv 1$, $f(\tilde{v})$ is smooth with a single maximum. The position of this maximum, which we will call $v_m$, shifts to the right as $A_1$ increases and becomes broader at the same time. This is as expected since $A_1$ governs the strength of positive noise pulses. Given our convention $a\geq 0$, the lowest value of $A_1$ we are allowing is fixed by $\lambda_1A_1=\lambda_2 A_2$, and gives the smallest $v_m$ for fixed $\lambda_1$, $\lambda_2$ and $A_2$. At that point the sign of $v_m$ is determined by the faster noise, i.e.\ $v_m>0$ for $\lambda_1>\lambda_2$.

The above trend, of $v_m$ increasing with $A_1$ and the peak broadening, holds also if only one of the rates is low ($<1$) but the other high ($>1$). In that case, however, the minimum is ``pinned'' so that it is always on the side of the faster noise. E.g.\ for $\lambda_1>1$, $\lambda_2<1$ one always has $v_m>0$.

Finally if both rates are low, the maximum of $f(\tilde{v})$ is at $v_m=0$. When also the total rate is low, i.e.\ $\lambda_1+\lambda_2<1$, the maximum turns into a power-law peak
\be
\label{f_power}
f(\tilde{v})\propto |\tilde{v}|^{\lambda_1+\lambda_2-1}
\ee
near $\tilde{v}=0$. This singular contribution is generically present also for larger rates but then has less dramatic effects. For example for $\lambda_1<1$, $\lambda_2<1$ but total noise rate $\lambda_1+\lambda_2>1$, the leading order decrease of $f(\tilde{v})$ from its maximum at $v_m=0$ follows the power law~(\ref{f_power}). In the case $\lambda_1=\lambda_2=1$, where the exponenet of this power law becomes one, an explicit form for $f(\tilde{v})$ can be found as
\be
f(\tilde{v})=\frac{1}{A_1^{-1}+A_2^{-1}}\left(\Theta(\tilde{v})e^{-\tilde{v}/A_1}+\Theta(-\tilde{v})e^{\tilde{v}/A_2}\right).
\ee
In agreement with the general discussion above this has a peak at $v_m=0$ and departs from this with terms proportional to $|\tilde{v}|$, leading to a jump discontinuity in the derivative at $\tilde{v}=0$.

Regarding the function $h(\tilde{v})$, we see directly from its definition in Eqs.~(\ref{LTsol}) and~(\ref{LTsol2}) that for small $\tilde{v}$ it will grow as $|\tilde{v}|^{\lambda_1+\lambda_2-1}$, thus matching the singular term~(\ref{f_power}) in $f(\tilde{v})$.  

In our further discussion we now need to distinguish $\Delta > a$ and $\Delta< a$.

\begin{figure}
\begin{tabular}{lr}
\includegraphics[height=3.3cm]{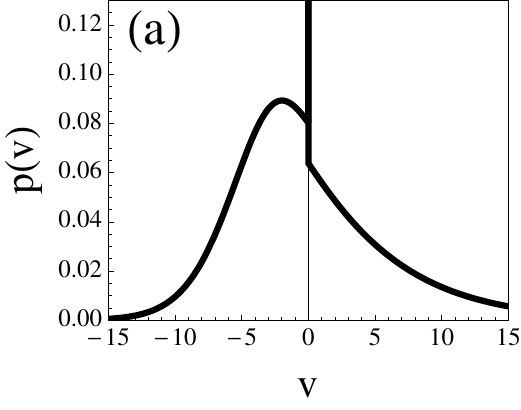}&\includegraphics[height=3.3cm]{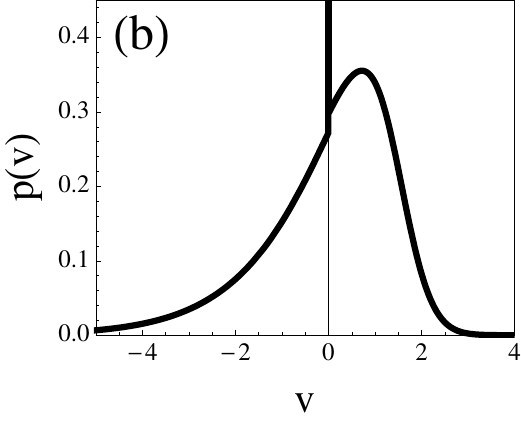}
\end{tabular}
\caption{\label{Fig_pdouble1} Characteristic double peaks in $p(v)$ for $\Delta>a$. Parameter values: (a) $\lambda_1=1.5$, $\lambda_2=7.5$, $A_1=5$, $\Delta=1$. (b) $\lambda_1=20$, $\lambda_2=2$, $A_1=0.1$, $\Delta=0.1$.}
\end{figure}

\subsection{The case $\Delta >a$}

For $\Delta >a$, $p(v)$ is given by Eq.~(\ref{pdist2}): the density consists of two distinct parts for $v>0$ and $v<0$, and exhibits a delta-peak at $v=0$ in addition to the discontinuity, as illustrated in Fig.~\ref{Fig_pdens1}. An additional maximum arises when the maximum in $p_+(v)$ (or $p_-(v)$), which is located at $v_m+v_+$ (or $v_m+v_-$, respectively), is ``visible'', i.e.\ is not cut-off by the discontinuity. The conditions for the appearance of these two possible maxima are thus
\be
\label{max1}
v_m+v_-=v_m+\Delta-a<0,
\ee
and
\be
\label{max2}
v_m+v_+=v_m-(\Delta+a)>0
\ee
where $v_m$ denotes the position of the maximum in $f(\tilde{v})$ as before. Since $v_->0$ and $v_+<0$, these conditions cannot both be satisfied, so at most one maximum can be present. We can then summarize the conditions as $\Delta < |a-v_m|$. Because we are looking at the regime $\Delta >a$, a maximum can then be visible in some range of $\Delta$ only if -- in $f(\tilde{v})$ -- it is at $v_m<0$ or at $v_m>2a$. This means in particular that the power law peak in $f(\tilde{v})$ that appears when $\lambda_1,\lambda_2<1$ is never visible as it is located at $v_m=0$.

Fig.~\ref{Fig_pdouble1} shows two plots of $p(v)$ for $\Delta>a$, which reveal the peak, the step discontinuity, and a maximum to both the left and the right of the peak. Whether the discontinuity at $v=0$ has a positive or negative sign is determined by the relative location of the maxima in $p_\pm(v)$, the skewness of $f(v)$, and the normalization constants $c_1,c_2$ and $\Gamma_0$ in Eq.~(\ref{pdist2}).

We close this subsection with a discussion of the physical intuition behind the results for the stationary velocity distribution $p(v)$. The regime considered here is the one of strong friction. The particle then experiences a restoring force for small velocity $v$ that changes sign at $v=0$. In the absence of a noise pulse, the particle velocity will therefore return to zero in finite time. With PSN, there is always a nonzero probability that the time interval between two noise pulses will be long enough for this to happen. Therefore one has a finite probability in steady state of finding the particle with zero velocity, as represented by the $\delta$-peak in $p(v)$. The secondary maximum in $p(v)$ is inherited from the maximum in $f(\tilde{v})$ and reflects the effects of positive and negative force pulses not having the same rates or amplitudes.

\subsection{The case $\Delta<a$}

\begin{figure}
\begin{center}
\includegraphics[width=8cm]{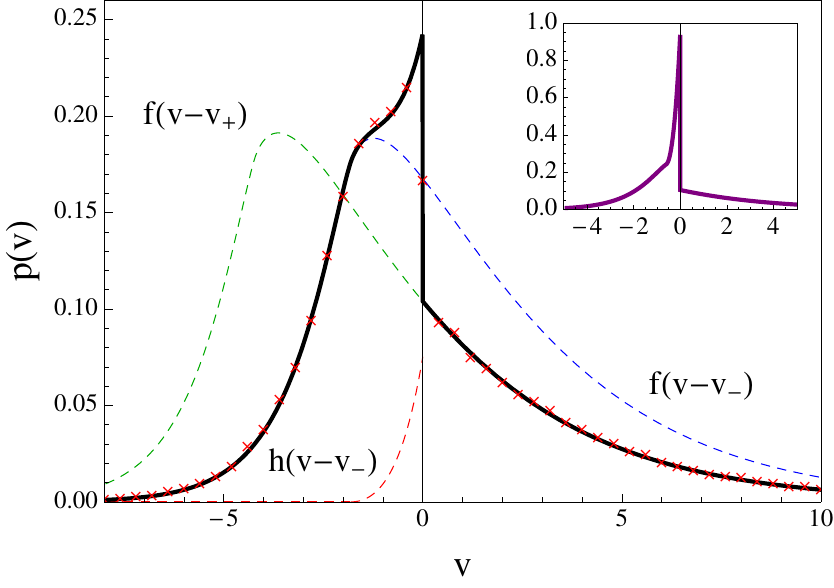}
\caption{\label{Fig_pdens2}(Colors online) The stationary density $p(v)$ for $\Delta<a$, Eq.~(\ref{pneg}), for 
parameter values $\lambda_1=\lambda_2=1.5$, $A_1=3$, $\Delta=0.4a=1.2$. The three distinct contributions to the density are indicated by dashed lines. Inset: $p(v)$ for larger friction parameter $\Delta=0.8a=2.4$.}
\end{center}
\end{figure}

In this case the density is given by Eq.~(\ref{pneg}): the delta-peak disappears, but $p(v)$ contains an additional contribution from the LT solution as indicated in Fig.~\ref{Fig_pdens2}. This contribution converges to the delta-peak as $\Delta\to a$ as illustrated in the inset of Fig.~\ref{Fig_pdens2}. Similarly to the regime $\Delta>a$, the density can contain an additional maximum. This requires that one of the conditions Eqs.~(\ref{max1},\ref{max2}) be satisfied, though that is not necessarily sufficient: the extra contribution $h(v-v_-)$ may cover up the maximum, as happens in Fig.~\ref{Fig_pdens2}.

\begin{figure}
\begin{tabular}{lr}
\includegraphics[height=3.3cm]{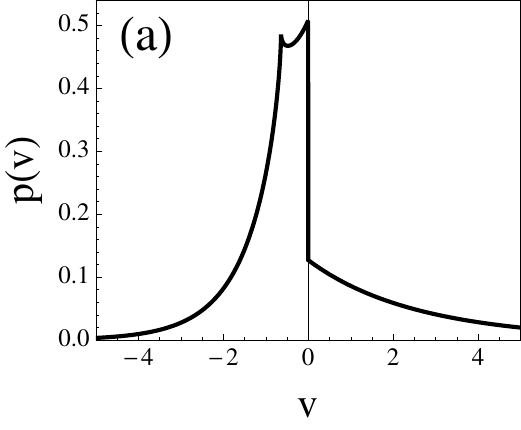}&\includegraphics[height=3.3cm]{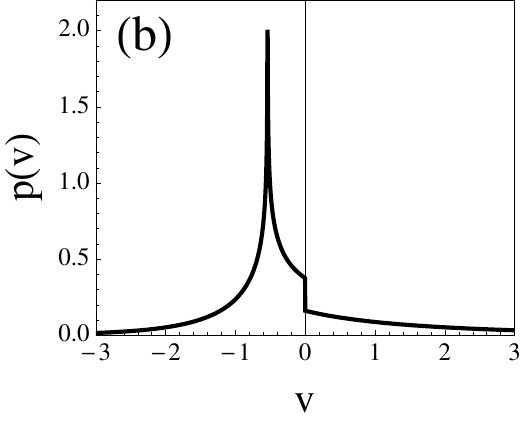}
\end{tabular}
\caption{\label{Fig_pdouble2}Characteristic peaks of the stationary density $p(v)$ for $\Delta<a$. Parameter values: (a) $\lambda_1=\lambda_2=0.8$, $A_1=3$, $\Delta=0.6a=0.96$. (b) $\lambda_1=\lambda_2=0.45$, $A_1=3$, $\Delta=0.4a=0.36$. }
\end{figure}

We find that a maximum in $p(v)$ for $v<0$ is generically visible for small $\Delta/a$, but as $\Delta$ becomes larger the LT contribution increasingly dominates and eventually makes $p(v)$ monotonically increasing again. In a narrow parameter range both the maximum inherited from $f(\tilde{v})$ and the increase towards $v=0$ from $h(v-v_-)$ are visible, giving two distinct peaks in $p(v)$, as shown in Fig.~\ref{Fig_pdouble2}a. For $\lambda_1+\lambda_2 \le 1$ the left of these peaks becomes a power law divergence with the power law $|v-v_-|^{\lambda_1+\lambda_2-1}$ from~(\ref{f_power}), cf.\ Fig.~\ref{Fig_pdouble2}b.

It is possible also to find parameter settings where a maximum in $p(v)$ is visible on the $v>0$ side, but this occurs only in a rather narrow parameter range. This maximum also never becomes singular as it is inherited from a maximum at $v_m>0$ (rather than $v_m=0$) in $f(\tilde{v})$.

For a physical interpretation of the results above, one needs to recall that in the absence of noise impulses the particle is subject to the force $F_\pm(v) = -(v-v_\pm)/\tau$ depending on the sign of $v$. As both $v_+$ and $v_-$ are negative for the weak friction case $\Delta<a$ considered here, the velocity of the particle will thus approach $v_-=-(a-\Delta)\tau$, exponentially in time. When noise rates are low there is then a high probability in steady state of finding the particle with velocity $v$ close to $v_-$, and this causes the peak at or near $v_-$ in $p(v)$.

\section{Transport properties}
\label{Sec_trans}

As discussed in the introduction, directed motion, i.e., a non-zero mean velocity of the object, results from the coupling of asymmetric fluctuations to a nonlinear response. If we focus on the stationary regime again, the mean velocity is given by
\be
\label{mv}
\langle v\rangle=\int_{-\infty}^\infty v\,p(v)\D v,
\ee
and is generally non-zero for $\lambda_1\neq \lambda_2$ or $A_1\neq A_2$. A simpler expression for $\langle v\rangle$ can be derived by taking directly the noise average in Eq.~(\ref{model}) and considering the steady state, where $\langle\dot{v}\rangle=0$. This yields
\be
\label{mv_simple}
\langle v\rangle&=&-\Delta\tau\langle\sigma(v)\rangle\nonumber\\
&=&\Delta\tau\left(\int_{-\infty}^0p(v)\D v-\int_0^\infty p(v)\D v\right).
\ee
Eq.~(\ref{mv_simple}) immediately tells us that a non-zero mean velocity results physically from the presence of \cite{Baule12}
\begin{enumerate}[(a)]
\item Inertia (non-zero $\tau$)
\item The influence of dry friction (non-zero $\Delta$)
\item An asymmetric stationary velocity distribution
\end{enumerate}
A positive mean velocity is induced when the total probability of observing a negative velocity is larger than the total probability of observing a positive one. This is somewhat counterintuitive, but is here a consequence of the particular form of $p(v)$. Since the friction is symmetric, the asymmetry in the stationary distribution can only be induced by an asymmetric noise $\xi(t)$. The asymmetric fluctuations are thus rectified by the nonlinear dry friction and lead to directed motion.

\begin{figure}
\begin{tabular}{lr}
\includegraphics[height=3.3cm]{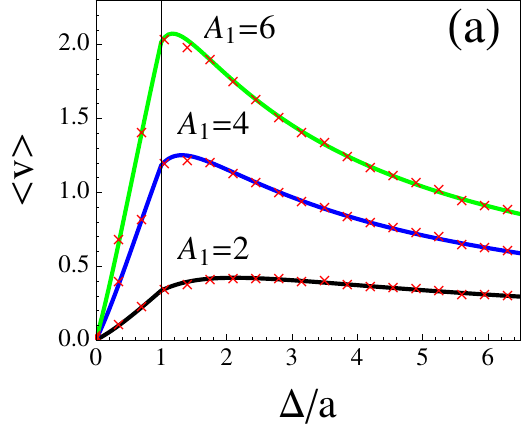}&\includegraphics[height=3.3cm]{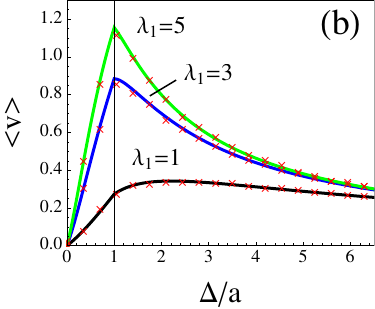}
\end{tabular}
\caption{\label{Fig_mvel}(Colors online) The mean velocity given by Eqs.~(\ref{mv_simple}) and (\ref{mv_simple2}). The cusp singularity at $\Delta=a$ is visible. (a) Plot of $\langle v\rangle$ for increasing values of $A_1$ ($\ge A_2$). Here $\lambda_1=\lambda_2=1.5$. (b) For larger $\lambda_1$ ($\ge\lambda_2$), the maximum in $\langle v\rangle$ shifts to $\Delta/a=1$. Here $\lambda_2=1$, $A_1=2$. }
\end{figure}

Note that Eq.~(\ref{mv_simple}) neglects any contribution from a delta-peak at $v=0$ and is thus valid only for $\Delta<a$. In order to obtain the equivalent expression for $\Delta>a$ one needs the average of $\sigma(v)$ over a delta-peak, which is a priori undefined because $\sigma(v)$ is discontinuous at $v=0$. To get a meaningful result one has to consider the linearized sign-function Eq.~(\ref{sigeps}). Integrating the linearized part over $[-\epsilon,\epsilon]$ using the distribution Eq.~(\ref{peps}) leads to a contribution
\be
\int_{-\epsilon}^\epsilon\sigma_\epsilon(v) p_\epsilon(v)\D v&\propto&\int_{-\epsilon}^\epsilon\frac{v}{\epsilon}\left|\frac{v}{\epsilon}+\frac{a}{\Delta} \right|^{\bar{\lambda}\epsilon/\Delta-1}\D v\nonumber\\
&=&-\frac{a}{\Delta}\int_{-\epsilon}^\epsilon \left|\frac{v}{\epsilon}+\frac{a}{\Delta} \right|^{\bar{\lambda}\epsilon/\Delta-1}\D v+\mathcal{O}(\epsilon).\nonumber\\
\ee
As $\epsilon\to 0$ the term of order $\epsilon$ vanishes, while the integral in the first term becomes the area of the delta-peak. Including the proper normalization, the contribution to the average of the sign-function from the delta peak in $p(v)$ can be read off as $-a\Gamma_0/\Delta$. For $\Delta>a$ the expression corresponding to Eq.~(\ref{mv_simple}) is then
\be
\label{mv_simple2}
\langle v\rangle&=&\Delta\tau\left(\int_{-\infty}^0p(v)\D v+\frac{a}{\Delta}\Gamma_0-\int_0^\infty p(v)\D v\right).
\ee

We are interested in the behaviour of the mean velocity as the strength of the dry friction is increased. Since we assume $a\ge 0$, the mean velocity is also always $\ge 0$. Clearly, for $\Delta=0$ we have $\langle v\rangle=0$ and transport is purely diffusive in this case. On the other hand, for $\Delta\gg a$ the dry friction dominates the dynamics, so that $\langle v\rangle\to 0$. Therefore, the mean velocity behaves in a non-monotonic way and reaches a maximum for a certain value of $\Delta/a$, which is shown in Fig.~\ref{Fig_mvel}. The exact location of the maximum is difficult to determine analytically due to the complicated form of the stationary distribution, but we can get a good qualitative picture by numerical evaluation.

The transport generally improves as $a$ increases. For $\lambda_2 A_2=\mbox{const}$, the maximum of the mean velocity becomes larger upon increasing either $\lambda_1$ or $A_1$ and its position shifts towards $\Delta/a=1$. This value indicates the change between the two distinct shapes of the stationary velocity distribution, discussed in Sec.~\ref{Sec_pdens}. In the plots of the mean velocity in Fig.~\ref{Fig_mvel} this transition manifests itself in a distinct cusp singularity at $\Delta=a$. Note that when varying $\lambda_1$ or $A_1$, the value of $a$ changes, so the curves in Fig.~\ref{Fig_mvel} are each for a different range of $\Delta$. For large $a$ optimal transport is achieved for the dry friction value $\Delta=a$. Physically, this optimal value can be understood by noting that $a$ acts as a constant negative force bias in the equation of motion (\ref{model}). For $\Delta=a$ this negative drift is cancelled exactly by the dry friction, so that motion induced by the fluctuations is predominantly in the forward direction. This transport behaviour is similar to the one observed for the special case of one-sided PSN, discussed in Ref.~\cite{Baule12}, since for larger $a$ the effect of the fluctuations in the backward direction becomes negligible. In the one-sided case, the critical friction strength $\Delta=a$ indicates a transition between two distinct modes of motion: {\it directed} motion for $0<\Delta<a$ and {\it unidirectional} motion for $\Delta\ge a$. In the unidirectional regime, the dry friction completely filters out the negative drift and no negative velocity fluctuations appear.  

For smaller $a$ the mean velocity exhibits a maximum for $\Delta>a$, since in addition to the negative drift the fluctuations in the backward direction are more relevant. The maximum is then also less pronounced. For larger $\Delta>a$ the dry friction again starts to increasingly inhibit the transport. Overall, our model therefore exhibits the striking effect that increasing dissipation enhances the transport in a wide range of parameter values.

\section{Discussion}
\label{Sec_dis}

We have investigated an exactly solvable nonlinear model for the directed motion of an object subject to dry friction and Poissonian shot-noise (PSN) with zero mean. The transport in this model behaves in a non-monotonic way as the dry friction strength is increased, exhibiting a maximum at a certain dry friction strength. The singular features of the stationary velocity distribution should also appear for more general piecewise-continuous forces subject to PSN. This follows from our discussion in Sec.~\ref{Sec_model}, where the discontinuity and the peak at $v=0$ are derived for a general piecewise force. However, the partial solutions $p_\pm(v)$ would be more difficult to obtain when the force branches $F_\pm(v)$ are each given by a nonlinear force. The same would be true if one wanted to go beyond a one-dimensional description. It is also not clear how non-integrable solutions would contribute for more general nonlinearities, given that there might be multiple velocity fixed points.

We were only able to solve the KF equation for noise amplitude distributions that are given as superpositions of single exponentials. In Fourier space, single exponentials lead to a Green's function in the form of a rational function. Inverting the stationarity condition Eq.~(\ref{zerocurr}) in Fourier space then leads to an ODE for the stationary distribution. Due to the local nature of the ODE, the solutions $p_\pm(v)$ can be obtained by considering the dynamics separately on either side of the discontinuity. On each side the force is linear and thus $p_\pm(v)$ can be derived in a straightforward way. These solutions are in fact identical to the ones obtained directly from the integral equation (\ref{zerocurr}) by restricting the force to $F_\pm(v)$ separately. However, this direct solution of Eq.~(\ref{zerocurr}) is not correct for arbitrary amplitude distributions due to the mixing of the dynamics across the discontinuity in Eq.~(\ref{zerocurr}).

Our results might be relevant for coarse-grained models of recently developed granular motors. These motors are in principle realizations of the classical Feynman ratchet and pawl system in a granular gas. In the experimental work of Ref.~\cite{Eshuis10}, a chiral rotor of millimeter-size is immersed in a gas of rigid particles. By coating the sides of the rotor with different materials, a ratchet effect can be realized, leading to directed rotations. On a coarse-grained level the influence of the granular gas at different particle densities, acting like random force impulses, could be modeled using PSN. This could take into account different spatial and temporal noise characteristics. Theoretical studies based on the Boltzmann equation including dry friction effects have already found singular features in the velocity distribution similar to the ones obtained here \cite{Talbot11}.

Our model could be implemented in a straightforward way in experimental setups similar to the ones used by Chaudhury {\em et al} \cite{Goohpattader09,Goohpattader10,Mettu10}. In fact, vibrations with PSN statistics should in principle be easier to implement than Gaussian noise, since the fluctuations appear on a separate timescale. Dry friction has recently been shown to be relevant in particular to biological soft matter systems, which opens up the possibility of realizing the transport properties of our model in a nanoscale system \cite{Yoshina03,Yoshina06,Wang09,Menzel10}. The thermal environment would provide an additional noise source, but overall the qualitative transport features should remain unchanged since thermal noise does not exert any net force. The key issue would be to find a suitable realization of the PSN, e.g.\ through intermittent chemical reactions.


\begin{thebibliography}{99}

\bibitem{Reimann02}P. Reimann, Phys. Rep. {\bf 361}, 57 (2002).

\bibitem{Haenggi09}P. H\"anggi and F. Marchesoni, Rev. Mod. Phys. {\bf 81}, 387 (2009).

\bibitem{Daniel02}S. Daniel and M. J. Chaudhury, Langmuir {\bf 18}, 3404 (2002).

\bibitem{Daniel04}S. Daniel, S. Sircar, J. Gliem, and M. J. Chaudhury, Langmuir {\bf 20}, 4085 (2004).

\bibitem{Mahadevan04}L. Mahadevan, S. Daniel, and M. K. Chaudhury, Proc. Nat. Ac. Sci. USA {\bf 101}, 23 (2004).

\bibitem{Daniel05}S. Daniel, M. K. Chaudhury, and P.-G. de Gennes, Langmuir {\bf 21}, 4240 (2005).

\bibitem{Eglin06}M. Eglin, M. A. Eriksson, and R. W. Carpick, App. Phys. Lett. {\bf 88}, 091913 (2006).

\bibitem{Buguin06}A. Buguin, F. Brochard, and P.-G. de Gennes, Eur. Phys. J. E {\bf 19}, 31 (2006).

\bibitem{Fleishman07}D. Fleishman, Y. Asscher, and M. Urbakh, J. Phys.: Cond. Matt. {\bf 19}, 096004 (2007).

\bibitem{Cebiroglu10}G. Cebiroglu, C. Weber, and L. Schimansky-Geier, Chem. Phys. {\bf 375}, 439 (2010).

\bibitem{Baule12}A. Baule and P. Sollich, EPL {\bf 97}, 20001 (2012).

\bibitem{Urbakh04}M. Urbakh, J. Klafter, D. Gourdon, and J. Israelachvili, Nature {\bf 430}, 525 (2004).

\bibitem{Persson}B. N. J. Persson, \textit{Sliding Friction: Physical Principles and Applications} (Springer, Berlin, 2000).

\bibitem{Feynman}R. P. Feynman and A. R. Hibbs, {\em Quantum Mechanics and Path Integrals} (McGraw-Hill, New York, 1965).

\bibitem{deGennes05}P.-G. de Gennes, J. Stat. Phys. {\bf 119}, 953 (2005).

\bibitem{Hayakawa04}H. Hayakawa, Physica D {\bf 205}, 48 (2005).

\bibitem{Baule10}A. Baule, E. G. D. Cohen, and H. Touchette, J. Phys. A: Math. Th. {\bf 43}, 025003 (2010).

\bibitem{Touchette10}H. Touchette, E. Van der Straeten, and W. Just, J. Phys. A: Math. Th. {\bf 43}, 445002 (2010).

\bibitem{Menzel10}A. M. Menzel and N. Goldenfeld, Phys. Rev. E {\bf 84}, 011122  (2011).

\bibitem{Baule11}A. Baule, H. Touchette, and E. G. D. Cohen, Nonlinearity {\bf 24}, 351 (2011).

\bibitem{Goohpattader09}P. S. Goohpattader, S. Mettu, and M. K. Chaudhury, Langmuir {\bf 25}, 9969 (2009).

\bibitem{Goohpattader10}P. S. Goohpattader and M. K. Chaudhury, J. Chem. Phys. {\bf 133}, 024702 (2010).

\bibitem{Mettu10}S. Mettu, and M. K. Chaudhury, Langmuir {\bf 26}, 8131 (2010).

\bibitem{Luczka95}J. Luczka, R. Bartussek, and P. H\"anggi, Europhys. Lett. {\bf 31}, 431 (1995).

\bibitem{Feller}W. Feller, {\em An Introduction to Probability Theory and Its Applications. Vol I \& II} (Wiley, New York, 1968).

\bibitem{Haenggi78}P. H\"anggi. Z. Phys. B {\bf 31}, 407 (1978).

\bibitem{Kim07}C. Kim, E.~K. Lee, P. HŠnggi, and P. Talkner, Phys. Rev. E {\bf 76}, 011109 (2007).

\bibitem{Eshuis10}P. Eshuis, K. van der Weele, D. Lohse, and D. van der Meer, Phys. Rev. Lett. {\bf 104}, 248001 (2010).

\bibitem{Talbot11}J. Talbot, R. D. Wildman, and P. Viot, Phys. Rev. Lett. {\bf 107}, 138001 (2011).

\bibitem{Yoshina03}C. Yoshina-Ishii and S. G. Bover, J. Am. Chem. Soc. {\bf 125}, 3696 (2003).

\bibitem{Yoshina06}C. Yoshina-Ishii, Y.-H. M. Chan, J. M. Johnson, L. A. Kung, P. Lenz, and S. G. Bover, Langmuir {\bf 22}, 5682 (2006).

\bibitem{Wang09}B. Wang, S. M. Anthony, S. C. Bae, and S. Granick, Proc. Nat. Acad. Sci. USA {\bf 106}, 15160 (2009).


\end{thebibliography}
\end{document}